\documentclass[12pt]{article}
\usepackage{amsthm,amsfonts,amsmath,amssymb,mathrsfs,url}
\usepackage[usenames,dvipsnames]{color}

\title{Multi-Time Formulation of Pair Creation}

\author{
S\"oren Petrat\footnote{Mathematisches Institut, 
	Ludwig-Maximilians-Universit\"at, Theresienstr. 39, 80333 M\"unchen, Germany.
	E-mail: petrat@math.lmu.de}\ \ and
Roderich Tumulka\footnote{Department of Mathematics,
     Rutgers University,
     110 Frelinghuysen Road, Piscataway, NJ 08854-8019, USA.
     E-mail: tumulka@math.rutgers.edu}
}
\date{January 23, 2014}

\theoremstyle{plain}
\theoremstyle{plain}
\theoremstyle{plain}
\theoremstyle{plain}
\theoremstyle{definition}

\newcommand{\be}{\begin{equation}}
\newcommand{\ee}{\end{equation}}
\newcommand{\Hilbert}{\mathscr{H}}

\newcommand{\RRR}{\mathbb{R}}

\newcommand{\CCC}{\mathbb{C}}

\newcommand{\scp}[2]{\langle #1|#2 \rangle}

\newcommand{\vy}{\boldsymbol{y}}
\newcommand{\valpha}{\boldsymbol{\alpha}}
\newcommand{\vx}{\boldsymbol{x}}

\newcommand{\sA}{\mathscr{A}}
\newcommand{\sB}{\mathscr{B}}

\newcommand{\sF}{\mathscr{F}}

\newcommand{\sM}{\mathscr{M}}

\newcommand{\sS}{\mathscr{S}}

\newcommand{\free}{\mathrm{free}}
\newcommand{\xbar}{\overline{x}}
\newcommand{\Mbar}{\overline{M}}
\newcommand{\vxbar}{\overline{\boldsymbol{x}}}
\newcommand{\rbar}{\overline{r}}
\newcommand{\ibar}{\overline\imath}
\newcommand{\jbar}{\overline\jmath}
\newcommand{\Gbar}{\overline{G}}
\newcommand{\abar}{\overline{a}}
\newcommand{\inter}{\mathrm{int}}
\newcommand{\Hd}{\mathcal{H}}

\begin{document}
\maketitle

\begin{abstract}
In a recent work \cite{pt:2013c}, we have described a formulation of a model quantum field theory in terms of a multi-time wave function and proposed a suitable system of multi-time Schr\"odinger equations governing the evolution of that wave function. Here, we provide further evidence that multi-time wave functions provide a viable formulation of relevant quantum field theories by describing a multi-time formulation, analogous to the one in \cite{pt:2013c}, of another model quantum field theory. This model involves three species of particles, say $x$-particles, anti-$x$-particles, and $y$-particles, and postulates that a $y$-particle can decay into a pair consisting of an $x$ and an anti-$x$ particle, and that an $x$--anti-$x$ pair, when they meet, annihilate each other creating a $y$-particle. (Alternatively, the model can also be interpreted as representing beta decay.) The wave function is a multi-time version of a time-dependent state vector in Fock space (or rather, the appropriate product of Fock spaces) in the particle-position representation. We write down multi-time Schr\"odinger equations and verify that they are consistent, provided that an even number of the three particle species involved are fermionic.

\medskip

\end{abstract}

\section{Introduction and Overview}

This note extends our discussion in \cite{pt:2013c} of multi-time wave functions in quantum field theory. There, we described a multi-time version of a model quantum field theory (QFT) involving two particle species, $x$-particles and $y$-particles, such that $x$-particles can emit and absorb $y$-particles, $x\rightleftarrows x+y$. Here, we set up a multi-time version of another model QFT involving three species, $x$-particles, $\xbar$-particles (which may be thought of as the anti-particles of the $x$-particles), and $y$-particles, with possible reactions $x + \xbar \rightleftarrows y$. The model is inspired by electrons ($x$), positrons ($\xbar$), and photons ($y$), but we do not try here to be as realistic as possible; rather, we aim at simplicity. In particular, we take all three species to be Dirac particles and do not exclude wave functions with contributions of negative energy; also, we ignore any connection between $x$-states of negative energy and $\xbar$-states. We also ignore the fact that our model Hamiltonian is ultraviolet divergent (already in the 1-time formulation) because that problem is orthogonal to the issue of setting up a multi-time formulation.

The set of multi-time equations that we propose and discuss in this paper, Equations \eqref{multi123} below, can be applied also to other physical situations besides pair creation. Since the equations are independent of whether or not $\xbar$ is the anti-particle of $x$, we may write the reaction $y\rightleftarrows x+\xbar$ more abstractly as $a\rightleftarrows b+c$. Each of the three species $a$, $b$, and $c$ can be chosen to be either fermions or bosons, and assigned an arbitrary mass. Thus, besides pair creation, another scenario that is included is, as a variation of the reaction $x\rightleftarrows x+y$ of \cite{pt:2013c}, the reaction $x\rightleftarrows z+y$ in which an $x$-particle emits a $y$-particle and mutates into a $z$-particle; conversely, a $z$ absorbing a $y$ becomes an $x$. For example, beta decay is of this type; in fact, beta decay (roughly speaking, the decay of a neutron into a proton, an electron, and an anti-neutrino) actually consists of two decay events, of which the first is
$d \to u+W^-$ (a down quark mutates into an up quark while emitting a negatively charged $W$ boson), and the second is $W^- \to e^- + \overline{\nu}_e$ (the $W^-$ decays into an electron and an anti-$\nu_e$, where $\nu_e$ is an electron-neutrino); that is, each of the two decay events is of the form $x\to z+y$. 

We remark that all fundamental processes of particle creation or annihilation in nature seem to be of one of the three forms $x\rightleftarrows x+y$, $x\rightleftarrows z+y$, or $y\rightleftarrows x+\xbar$. Our work in \cite{pt:2013c} and in the present paper shows that all of these processes can be formulated in terms of multi-time equations. Thereby, it contributes evidence in favor of the hypothesis that multi-time wave functions provide a viable formulation of all relevant QFTs.

Multi-time wave functions were considered early on in the history of quantum theory, also for quantum field theory \cite{dirac:1932,dfp:1932,bloch:1934}. However, in these early papers, only electrons were regarded as particles, and photons were replaced by a field configuration. Bloch~\cite{bloch:1934} noted that multi-time equations require consistency conditions (see also \cite{pt:2013a}). Multi-time wave functions with a variable number of time variables (see below for explanation) were considered before in \cite{schweber:1961,DV82b,DV85,Nik10,pt:2013c}. Droz-Vincent \cite{DV82b,DV85} gave an example of a consistent multi-time evolution with interaction which, however, does not correspond to any ordinary QFT model. For further references about multi-time wave functions, a deeper discussion of consistency conditions, and no-go results about interaction potentials in multi-time equations, see \cite{pt:2013a}. For a comparison of the status and significance of multi-time formulations in classical and quantum physics, see \cite{pt:2013e}.

The plan of this paper is as follows. After introducing notation in Section~\ref{sec:notation}, we first describe the model QFT in Section~\ref{sec:1time} in the usual one-time formulation and then give a multi-time formulation in Section~\ref{sec:multitime}; in particular, we specify a system of multi-time Schr\"odinger equations\footnote{The expression \emph{Schr\"o\-din\-ger equation} is not meant to imply that the Hamiltonian involves the Laplace operator, but is understood as including the Dirac equation.} for the multi-time wave function $\phi$, see \eqref{multi123}. The function $\phi$ is defined on the set $\sS_{x\xbar y}$ of \emph{spacelike configurations}, which means any finite number of space-time points that are mutually spacelike or equal,\footnote{Note our usage of  the word \emph{spacelike}: Two space-time points that are mutually spacelike are unequal, but a spacelike configuration may contain two particles at the same space-time point.} with each point marked as either an $x$-, an $\xbar$-, or a $y$-particle. The set $\sS_{x\xbar y}$ can be regarded as a subset of $\Gamma(\RRR^4)\times \Gamma(\RRR^4)\times \Gamma(\RRR^4)$, with one factor for each particle species, $\RRR^4$ representing space-time, and the notation
\be\label{Gammadef}
\Gamma(S) = \bigcup_{N=0}^\infty S^N
\ee
for any set $S$.
Since a function $f$ on $\Gamma(S)$ can be thought of as consisting of one function $f^{(N)}$ on each sector $S^N$, and since $f^{(N)}$ is a function of $N$ variables with values in $S$, we say that $f$ is a function with a variable number of variables. Correspondingly, $\phi$ on $\sS_{x\xbar y}$ involves a variable number of time variables. If all time variables in $\phi$ are set equal (relative to some Lorentz frame $L$), we recover the wave function $\psi$ of the 1-time formulation. We then verify the consistency of our proposed system of equations in Section~\ref{sec:consistency}; that is, we verify (on a non-rigorous level) that the system possesses a unique solution $\phi$ for every initial datum $\phi_0$ (at all times set equal to zero). Our proposed multi-time QFT model is not fully covariant because it involves, as a coefficient of the pair creation term, an arbitrary but fixed complex-linear mapping $\tilde{g}:S\to S\otimes S$, with $S=\CCC^4$ the 4-dimensional complex spin space used in the Dirac equation, and nonzero mappings $\tilde{g}$ of this type are never Lorentz-invariant. Apart from this point, the model is covariant, as can be seen from the Lorentz transformation of the multi-time Schr\"odinger equations that we carry out in Section~\ref{sec:lorentz}. The natural generalization to curved space-time is described in Section~\ref{sec:curved}.

As a last remark in this section, we find that the multi-time equations of the reaction $a\rightleftarrows b+c$ are consistent if and only if an \emph{even} number of the three species $a,b,c$ are \emph{fermions}, and an \emph{odd} number of them are \emph{bosons} (also in case some of the species coincide, such as $a=b$). That is, either all three are bosons, or two of them are fermions and one is a boson. This rule for which kinds of reactions are allowed agrees with which kinds of reactions occur in nature.\footnote{Indeed, a fermion cannot decay into two fermions, nor into two bosons, whereas a boson can decay into two fermions (e.g., photon $\to$ electron + positron) or two bosons (e.g., Higgs $\to W^+ + W^-$); furthermore, a fermion can decay into a fermion and a boson (e.g., $d\to u+W^-$) but a boson cannot.} This rule can also be derived from the conservation of spin and the spin--statistics relation (i.e., that bosons have integer and fermions half-odd spin), but here we obtain it differently, from the consistency of the multi-time equations. That is, a single-time formulation (i.e., a Hilbert space and Hamiltonian) can be (consistently) specified for each of the 8 combinations of $a,b,c$ being bosons or fermions, but in the multi-time formulation the rule above restricts the possibilities, thereby providing a possible explanation of why reactions violating the rule do not occur in nature.

\section{Notation}
\label{sec:notation}

Our notation follows that of \cite{pt:2013c}. Space-time points are denoted by $x=(x^0,\vx)$ etc.; vectors in 3-dimensional physical space $\RRR^3$ are denoted by bold symbols such as $\vx$. When writing down the time evolution equations, we make very many details explicit, at the risk of making the equations appear more complex than they are, in order to ensure full clarity. A spacelike configuration $q^4\in\sS_{x\xbar y}$ will mostly be written as $q^4=\bigl( x^{4M},\xbar^{4\Mbar}, y^{4N}\bigr)$ with the $x$-configuration $x^{4M} = (x_1,\ldots,x_j, \ldots,x_M)$, the $\xbar$-configuration $\xbar^{4\Mbar} = (\xbar_1,\ldots,\xbar_{\jbar}, \ldots,\xbar_{\Mbar})$, and the $y$-configuration $y^{4N} = (y_1,\ldots,y_k, \ldots,y_N)$. All spin indices refer to the spin space $S=\CCC^4$ of the Dirac equation and thus run from 1 to 4; $r_j$ is the spin index of particle $x_j$, $\rbar_{\jbar}$ that of $\xbar_{\jbar}$, and $s_k$ that of $y_k$. We will often write only those spin indices that are of particular interest in the expression at hand; e.g., we may write $\psi_{s_{N+1}}$ for a wave function that still has the usual indices $r_1,\ldots, s_N$ but also one additional index $s_{N+1}$ (that is understood as listed after $s_N$). A hat (\,$\widehat{\ }$\,) denotes omission; e.g., $\psi_{\widehat{s_k}}$ is the wave function with the usual indices except $s_k$.

Configurations on the hyperplane $t=0$ (relative to some Lorentz frame $L$) or just in 3-dimensional space, $q^3\in \Gamma(\RRR^3)^3$, are denoted by $q^3=\bigl( x^{3M},\xbar^{3\Mbar}, y^{3N}\bigr)$ with the $x$-configuration $x^{3M} = (\vx_1,\ldots,\vx_j, \ldots,\vx_M)$, the $\xbar$-configuration $\xbar^{3\Mbar} = (\vxbar_1,\ldots,\vxbar_{\jbar}, \ldots,\vxbar_{\Mbar})$, and the $y$-configuration $y^{3N} = (\vy_1,\ldots,\vy_k, \ldots,\vy_N)$. The notation $x^{3M}\setminus \vx_j$ means that $\vx_j$ is omitted, $x^{3M}\setminus \vx_j = (\vx_1,\ldots,\vx_{j-1},\vx_{j+1},\ldots,\vx_M)$; likewise with $x^{4M}\setminus x_j$.

\section{Single-Time Formulation}
\label{sec:1time}

The model that this paper is about is inspired by the models in chapter 12 of \cite{schweber:1961}. Let $\varepsilon_x$ be $+1$ if the $x$-particles are bosons, and $-1$ if they are fermions; likewise with $\varepsilon_{\xbar}$ and $\varepsilon_y$. The 1-time wave function $\psi(t,q^3)$ is, at any time $t\in\RRR$, a spinor-valued function on $\Gamma(\RRR^3)^3$,
\be
\psi=\psi\bigl(x^{3M},\xbar^{3\Mbar},y^{3N}\bigr)=\psi_{r_1\ldots r_M\rbar_1\ldots \rbar_{\Mbar}s_1\ldots s_N}\bigl(x^{3M},\xbar^{3\Mbar},y^{3N}\bigr)\,,
\ee
with the appropriate symmetry,
\begin{subequations}\label{sym}
\begin{align}
\psi_{r_ir_j}(\ldots, x_i, \ldots ,x_j, \ldots) &= \varepsilon_x \, \psi_{r_jr_i}(\ldots, x_j, \ldots, x_i,\ldots)\,,\label{xsym}\\
\psi_{\rbar_{\ibar}\rbar_{\jbar}}(\ldots, \xbar_{\ibar}, \ldots ,\xbar_{\jbar}, \ldots) &= \varepsilon_{\xbar} \, \psi_{\rbar_{\jbar}\rbar_{\ibar}}(\ldots, \xbar_{\jbar}, \ldots, \xbar_{\ibar},\ldots)\,,\label{xbarsym}\\
\psi_{s_ks_\ell}(\ldots, y_k, \ldots ,y_\ell, \ldots) &= \varepsilon_y \, \psi_{s_\ell s_k}(\ldots, y_\ell, \ldots, y_k,\ldots)\,,\label{ysym}
\end{align}
\end{subequations}
where the dots signify that all other variables are unchanged, and unchanged spin indices were not written at all (so $\psi_{s_\ell s_k}$ means that indices $s_k$ and $s_\ell$ have been interchanged). Correspondingly, $\psi$ lies in the Hilbert space
\be\label{Hilbertdef}
\Hilbert=\sF_x\otimes \sF_{\xbar}\otimes \sF_y
\ee
with 
\be\label{Fockdef}
\sF_u = \bigoplus_{N=0}^\infty S_{\varepsilon_u} L^2\Bigl(\RRR^{3N},(\CCC^4)^{\otimes N}\Bigr)
\ee
the Fock space for species $u\in\{x,\xbar,y\}$ and $S_{+1},S_{-1}$ the symmetrization and anti-symmetrization operators, respectively. The inner product in $\Hilbert$ is
\begin{multline}
\scp{\psi}{\chi} = \sum_{M,\Mbar,N=0}^\infty \:\int\limits_{\RRR^{3M}} dx^{3M} \!\! \int\limits_{\RRR^{3\Mbar}}d\xbar^{3\Mbar}
\!\! \int\limits_{\RRR^{3N}}dy^{3N} \!\! \sum_{r_1\ldots r_M=1}^4 \sum_{\rbar_1\ldots \rbar_{\Mbar}=1}^4\sum_{s_1\ldots s_N=1}^4 \times\\
\times\: \psi^*_{r_1\ldots s_N}\bigl(x^{3M},\xbar^{3\Mbar},y^{3N}\bigr) \, \chi_{r_1\ldots s_N}\bigl(x^{3M},\xbar^{3\Mbar},y^{3N}\bigr)\,.
\end{multline}

The 1-time wave function $\psi$ evolves according to the 1-time Schr\"odinger equation ($\hbar=1$)
\begin{subequations}\label{Schr1Hdef}
\be\label{Schr1}
i\frac{\partial \psi}{\partial t}=H\psi
\ee
with the 1-time Hamiltonian $H$ on $\Hilbert$ given by
\begin{align}\label{Hdef}
&H\psi\bigl(x^{3M},\xbar^{3\Mbar},y^{3N}\bigr) = \sum_{j=1}^MH_{x_j}^\free \psi + \sum_{\jbar=1}^{\Mbar}H_{\xbar_{\jbar}}^\free \psi + \sum_{k=1}^N H_{y_k}^\free\psi\:+\nonumber\\[3mm]
&\quad+ \sqrt{\frac{N+1}{M\Mbar}} \sum_{j=1}^M \sum_{\jbar=1}^{\Mbar} \varepsilon_x^{j+1} \, \varepsilon_{\xbar}^{\jbar+1}\, \varepsilon_y^N
 \sum_{s_{N+1}=1}^4 g_{r_j\rbar_{\jbar}\, s_{N+1}}\delta^3(\vx_j-\vxbar_{\jbar}) \,
    \times\nonumber\\&\qquad\qquad \times\: 
\psi_{\widehat{r_j}\widehat{\rbar_{\jbar}}\, s_{N+1}}\Bigl(x^{3M}\setminus \vx_j, \xbar^{3\Mbar} \setminus \vxbar_{\jbar}, (y^{3N}, \vx_j)\Bigr)\nonumber\\[4mm]
&\quad+ \sqrt{\frac{(M+1)(\Mbar+1)}{N}} \sum_{k=1}^N \varepsilon_x^M\,\varepsilon_{\xbar}^{\Mbar}\,\varepsilon_y^{k+1} 
\sum_{r_{M+1}\rbar_{\Mbar+1}=1}^4 g^*_{r_{M+1}\rbar_{\Mbar+1} s_k}\,
    \times\nonumber\\&\qquad\qquad \times\: 
\psi_{r_{M+1} \rbar_{\Mbar+1}\widehat{s_k}}\Bigl((x^{3M}, \vy_k),(\xbar^{3\Mbar},\vy_k),y^{3N}\setminus \vy_k\Bigr)\,,
\end{align}
\end{subequations}
where the term in the second and third line is meant to vanish whenever $M=0$ or $\Mbar=0$, the term in the fourth and fifth line is meant to vanish whenever $N=0$, $\varepsilon=\pm 1$ (as readers may recall) depending on whether the particles are bosons or fermions, and the free Hamiltonians are free Dirac operators (with speed of light set to $c=1$),
\begin{subequations}\label{Hfreedef}
\begin{align}
H^\free_{x_j} \psi_{r_j}(x^{3M},\xbar^{3\Mbar},y^{3N}) &= \sum_{r_j'=1}^4\biggl( -i \sum_{a=1}^3  (\alpha_a)_{r_jr_j'} \, \frac{\partial}{\partial x_j^a} + m_x  \beta_{r_jr_j'} \biggr) \psi_{r_j'}(x^{3M},\xbar^{3\Mbar},y^{3N}) \\
H^\free_{\xbar_{\jbar}} \psi_{\rbar_{\jbar}}(x^{3M},\xbar^{3\Mbar},y^{3N}) &= \sum_{\rbar_{\jbar}'=1}^4\biggl( -i \sum_{a=1}^3  (\alpha_a)_{\rbar_{\jbar}\rbar_{\jbar}'} \, \frac{\partial}{\partial \xbar_{\jbar}^a} + m_{\xbar}  \beta_{\rbar_{\jbar}\rbar_{\jbar}'} \biggr) \psi_{\rbar_{\jbar}'}(x^{3M},\xbar^{3\Mbar},y^{3N}) \\
H^\free_{y_k} \psi_{s_k}(x^{3M},\xbar^{3\Mbar},y^{3N}) &= \sum_{s_k'=1}^4\biggl(\! -i \sum_{a=1}^3  (\alpha_a)_{s_ks_k'} \, \frac{\partial}{\partial y_k^a} + m_y  \beta_{s_ks_k'} \!\biggr) \psi_{s_k'}(x^{3M},\xbar^{3\Mbar},y^{3N})
\end{align}
\end{subequations}
with mass parameters $m_x,m_{\xbar},m_y\geq 0$.

Equivalently, the Hamiltonian can be written as
\be
H=H_x + H_{\xbar} + H_y + H_{\inter}
\ee
with
\begin{subequations}
\begin{align}
H_{x} &= \int d^3\vx \sum_{r,r'=1}^4 a_r^\dagger(\vx) \bigl(-i\valpha_{rr'}\cdot \nabla+m_x\beta_{rr'}\bigr) a_{r'}(\vx)\label{Hxa}\\
H_\inter &=  \int d^3\vx\sum_{r,\rbar,s=1}^4 \Big( g_{r\rbar s} \, a_r^\dagger(\vx)\, \abar_{\rbar}^\dagger(\vx)\, b_s(\vx) + g_{r\rbar s}^* \, a_r(\vx)\, \abar_{\rbar}(\vx)\, b_s^\dagger(\vx) \Big)
\end{align}
\end{subequations}
and expressions analogous to \eqref{Hxa} for $H_{\xbar}$ and $H_y$. Here, $^\dagger$ denotes the adjoint operator, and $a_s(\vx),\abar_s(\vx),b_s(\vx)$ the annihilation operators for an $x,\xbar,y$-particle with spin component $s$ at location $\vx$ in position space, explicitly defined by
\begin{subequations}
\begin{align}
\bigl(a_r(\vx)\,\psi\bigr)(x^{3M},\xbar^{3\Mbar},y^{3N}) 
&= \sqrt{M+1}\; \varepsilon_x^M\, \psi_{r_{M+1}=r}\bigl((x^{3M},\vx),\xbar^{3\Mbar},y^{3N}\bigr)\,,\label{adef}\\
\bigl(a_r^\dagger(\vx)\,\psi\bigr)(x^{3M},\xbar^{3\Mbar},y^{3N}) 
&= \frac{1}{\sqrt{M}} \sum_{j=1}^M \varepsilon_x^{j+1} \, \delta_{rr_j}\,\delta^3(\vx_j-\vx)\, \psi_{\widehat{r_j}}\bigl(x^{3M}\setminus \vx_j,\xbar^{3\Mbar},y^{3N}\bigr)
\end{align}
\end{subequations}
and correspondingly for $\abar_s(\vx)$ and $b_s(\vx)$.

\section{Multi-Time Formulation}
\label{sec:multitime}

The constant $\kappa\in \RRR$ that we use below can be chosen to be $1/2$; we will show in Remark~\ref{rem:kappa} below that its value actually does not matter. The multi-time wave function $\phi$ is supposed to be defined on $\sS_{x\xbar y}$ and have values $\phi(x^{4M},\xbar^{4\Mbar},y^{4N})\in S^{\otimes (M+ \Mbar+N)}$ with $S=\CCC^4$; it is governed by the multi-time Schr\"odinger equations
\begin{subequations}\label{multi123}
\begin{align}
&i\frac{\partial}{\partial x_j^0}\phi\bigl(x^{4M},\xbar^{4\Mbar},y^{4N}\bigr) = H_{x_j}^\free\phi\bigl(x^{4M},\xbar^{4\Mbar},y^{4N}\bigr)
+ \kappa \sqrt{\frac{N+1}{M\Mbar}} \sum_{\jbar=1}^{\Mbar} \varepsilon_x^{j+1}\,\varepsilon_{\xbar}^{\jbar+1}\,\varepsilon_y^N  \,\times\nonumber\\
&\qquad \times  \sum_{s_{N+1}=1}^4 
\Gbar_{r_j \,\rbar_{\jbar}\, s_{N+1}}(\xbar_{\jbar}-x_j)\, 
\phi_{\widehat{r_j}\,\widehat{\rbar_{\jbar}}\,s_{N+1}}\Bigl(x^{4M}\setminus x_j,\xbar^{4\Mbar}\setminus \xbar_{\jbar}, (y^{4N}, x_j)\Bigr)\label{multi1}\\[3mm]
&i\frac{\partial}{\partial \xbar_{\jbar}^0}\phi\bigl(x^{4M},\xbar^{4\Mbar},y^{4N}\bigr) = H_{\xbar_{\jbar}}^\free \phi\bigl(x^{4M},\xbar^{4\Mbar},y^{4N}\bigr) 
+ (1-\kappa) \sqrt{\frac{N+1}{M\Mbar}}\sum_{j=1}^M \varepsilon_x^{j+1}\,\varepsilon_{\xbar}^{\jbar+1}\,\varepsilon_y^N\, \times\nonumber\\
&\qquad\times \: \sum_{s_{N+1}=1}^4 
G_{r_j \, \rbar_{\jbar}\, s_{N+1}}(x_j-\xbar_{\jbar})\, \phi_{\widehat{r_j}\,\widehat{\rbar_{\jbar}}\,s_{N+1}}\Bigl(x^{4M}\setminus x_j, \xbar^{4\Mbar}\setminus \xbar_{\jbar}, (y^{4N}, \xbar_{\jbar})\Bigr)\label{multi2}\\[3mm]
&i\frac{\partial}{\partial y_k^0}\phi\bigl(x^{4M},\xbar^{4\Mbar},y^{4N}\bigr) = H_{y_k}^\free\phi\bigl(x^{4M},\xbar^{4\Mbar},y^{4N}\bigr) 
+ \sqrt{\frac{(M+1)(\Mbar+1)}{N}} \varepsilon_x^M\,\varepsilon_{\xbar}^{\Mbar}\,\varepsilon_y^{k+1} \,\times\nonumber\\
&\qquad\times \: \sum_{r_{M+1},\rbar_{\Mbar+1}=1}^4
g^*_{r_{M+1} \rbar_{\Mbar+1} s_k} \, \phi_{r_{M+1} \,\rbar_{\Mbar+1} \widehat{s_k}}\Bigl( (x^{4M}, y_k),(\xbar^{4\Mbar}, y_k),y^{4N}\setminus y_k \Bigr)\,,\label{multi3}
\end{align}
\end{subequations}
where $G_{r\rbar s}(x)$ and $\Gbar_{r\rbar s}(\xbar)$ are appropriate Green functions, defined to be the solutions of the free Dirac equations
\begin{subequations}
\begin{align}
i\frac{\partial G}{\partial x^0} &= H_x^\free G\label{Gevol}\\[2mm]
i\frac{\partial \Gbar}{\partial \xbar^0} &= H_{\xbar}^\free \Gbar\label{Gbarevol}
\end{align}
with initial conditions
\begin{align}
G_{r\rbar s}(0,\vx) &= g_{r\rbar s} \, \delta^3(\vx)\\
\Gbar_{r\rbar s}(0,\vxbar) &= g_{r\rbar s} \, \delta^3(\vxbar)\,.
\end{align}
\end{subequations}
We will show in Section~\ref{sec:consistency} that the system \eqref{multi123} is consistent on $\sS_{x\xbar y}$ if and only if
\be\label{productepsilons}
\varepsilon_x \, \varepsilon_{\xbar} \, \varepsilon_y =1\,,
\ee
i.e., if and only if the number of fermionic species is even and that of bosonic ones is odd.

\bigskip

\noindent{\bf Remarks.}
\begin{enumerate}
\item The single-time evolution \eqref{Schr1Hdef} is contained in the multi-time equations \eqref{multi123} (for any $\kappa$) in the sense that if we set all time variables equal in $\phi$ then the 1-time wave function thus obtained, $\psi(t,x^{3M},\xbar^{3\Mbar},y^N)=\phi(t,\vx_1,\ldots,t,\vy_N)$, obeys \eqref{Schr1Hdef}. 

\item\label{rem:collision} As discussed in Remark 5 in Section 2.2 of \cite{pt:2013c}, something needs to be said about how \eqref{multi123} should be understood at the tips of $\sS_{x\xbar y}$ (i.e., at configurations with two particles at the same location, henceforth called \emph{collision configurations}). That is because, for example at a configuration with $x_j=y_k$, $\partial \phi/\partial x_j^0$ does not make sense for a function $\phi$ on $\sS_{x\xbar y}$ as varying $x^0_j$ while keeping $y_k$ fixed will lead away from $\sS_{x\xbar y}$. The rule formulated in \cite{pt:2013c} for these cases should be adopted here as well: At such a configuration, \eqref{multi123} should be understood as prescribing the directional derivative normally written as $(\partial/\partial x_j^0 + \partial/\partial y_k^0)\phi$ (and correspondingly for other collisions).

\item\label{rem:kappa} The system \eqref{multi123} for any value of $\kappa\in\RRR$ is actually equivalent to the system \eqref{multi123} for any other value of $\kappa\in\RRR$. Indeed, it is known \cite[Thm.~1.2 on p.~15]{thaller:1992} that the Green function of the Dirac equation (such as $G$ and $\Gbar$) vanishes on spacelike vectors; thus, the terms involving $\kappa$ vanish at \emph{collision-free} spacelike configurations. They do not vanish at a collision between (say) $x_j$ and $\xbar_{\jbar}$; there, according to Remark~\ref{rem:collision} above, the equations \eqref{multi123} are understood as specifying not $\partial \phi/\partial x_j^0$ and $\partial \phi/\partial \xbar_{\jbar}^0$ individually but only their sum, from which the factor $\kappa$ cancels out.

Thus, the solution $\phi$ of \eqref{multi123} is independent of $\kappa$; put differently, the one multi-time evolution law possesses several representations by systems of equations (involving different values of $\kappa$), roughly analogous to the way a gauge field possesses several representations corresponding to different gauges. That is, the choice of $\kappa$ is a matter of mere aesthetics; natural choices are $\kappa=0$, $\kappa=1/2$, or $\kappa=1$.

\item In the same way as for Assertion~3 in \cite{pt:2013c}, one can show that the multi-time wave function $\phi$ of \eqref{multi123} can be expressed as follows in terms of the particle annihilation operators at any $(x^{4M},\xbar^{4\Mbar},y^{4N})\in \sS_{x\xbar y}$:
\begin{multline}
\phi\bigl(x^{4M},\xbar^{4\Mbar},y^{4N}\bigr) = \frac{\varepsilon_x^{M(M-1)/2}
\varepsilon_{\xbar}^{\Mbar(\Mbar-1)/2} \varepsilon_y^{N(N-1)/2}}{\sqrt{M!\Mbar!N!}} \:\times\\
\times \: \Bigl\langle \emptyset \Big| a_{r_1}(x_1)\cdots a_{r_M}(x_M)\, \abar_{\rbar_1}(\xbar_1) \cdots \abar_{\rbar_{\Mbar}}(\xbar_{\Mbar})\, b_{s_1}(y_1)\cdots b_{s_N}(y_N)  \Big| \psi_0 \Bigr\rangle\,.
\end{multline}
Here, $\psi_0$ is obtained from $\phi$ by setting all time variables equal to zero, and $a_r(t,\vx)=e^{iHt}a_r(\vx)e^{-iHt}$ is the Heisenberg-evolved annihilation operator with the single-time Hamiltonian $H$ as in \eqref{Hdef}.

\item For any spacelike hypersurface $\Sigma$, let $\psi_\Sigma$ be the restriction of $\phi$ to $\Gamma(\Sigma)^3$ (i.e., to configurations on $\Sigma$), and let $\tilde\psi_\Sigma=F_{\Sigma\to\Sigma_0}\psi_\Sigma$ be the interaction picture wave function, where $\Sigma_0$ is the hypersurface of $t=0$ (in the Lorentz frame $L$) and $F_{\Sigma_1\to\Sigma_2}$ is the free time evolution from $\Sigma_1$ to $\Sigma_2$. Then $\tilde\psi$ obeys the Tomonaga--Schwinger equation
\be\label{TS}
i\bigl(\tilde\psi_{\Sigma'}-\tilde\psi_\Sigma\bigr) = \biggl( \int_{\Sigma}^{\Sigma'} \!\!\!\! d^4 x \, \Hd_I(x)\biggr)\, \tilde\psi_\Sigma
\ee
for infinitesimally neighboring spacelike hypersurfaces $\Sigma,\Sigma'$, with the interaction Hamiltonian density in the interaction picture given by
\be
\Hd_I(t,\vx)= e^{iH^\free t} 
\sum_{r,\rbar,s=1}^4 \Bigl( g_{r\rbar s} \, a_r^\dagger(\vx)\, \abar_{\rbar}^\dagger(\vx)\, b_s(\vx) + g_{r\rbar s}^* \, a_r(\vx)\, \abar_{\rbar}(\vx)\, b_s^\dagger(\vx) \Bigr) e^{-iH^\free t}\,.
\ee
This can be derived in much the same way as Assertion~5 in \cite{pt:2013c}, using Assertion~10 of \cite{pt:2013c}.
\end{enumerate}

\section{Consistency}
\label{sec:consistency}

As discussed in detail in \cite{pt:2013a}, there is a non-trivial condition for a system of multi-time equations to be consistent, i.e., to possess a solution $\phi$ for arbitrary initial conditions at time 0 (i.e., all times equal to 0). In Sections~5.1--5.5 of \cite{pt:2013c}, we have shown for multi-time equations of the type considered there that (leaving aside the ultraviolet divergence) they are consistent on the set of spacelike configurations if, at every spacelike configuration $q$, the \emph{consistency condition}
\be\label{consistency}
\biggl[ i\frac{\partial}{\partial t_j}-H_{j}, i\frac{\partial}{\partial t_k}-H_{k}\biggr]=0
\ee
holds for any two \emph{non-colliding} particles $j,k$ belonging to $q$ (see in particular Footnote 13 in Section 5.5 of \cite{pt:2013c}). The arguments described there apply to \eqref{multi123} as well. Therefore, the consistency of \eqref{multi123} follows if \eqref{consistency} holds for any two non-colliding particles, where $H_j\phi$ is the right-hand side of \eqref{multi1}, \eqref{multi2}, or \eqref{multi3}, whichever is appropriate depending on the species of particle $j$.

This is indeed the case if \eqref{productepsilons} holds. We report here the commutators. Any pair of particles is of one of these six forms: $xx$, $\xbar\xbar$, $yy$, $x\xbar$, $xy$, or $\xbar y$. Thus, six commutators need to be computed. Let us begin with the $yy$ commutator (and note that $N\geq 2$ whenever there are two $y$ variables for which we can consider the commutator):
\begin{align}
\label{comm3_neu}&\Big[i\partial_{y_k^0} - H_{y_k}, i\partial_{y_{\ell}^0} - H_{y_{\ell}}\Big] \phi \nonumber \\ 
&= \sqrt{\frac{(M+1)(M+2)(\Mbar+1)(\Mbar+2)}{N(N-1)}} \sum_{r_{M+1},r_{M+2},\rbar_{\Mbar+1},\rbar_{\Mbar+2}=1}^4 g^*_{r_{M+1}\rbar_{\Mbar+1}s_k} g^*_{r_{M+2}\rbar_{\Mbar+2}s_{\ell}} \times \nonumber \\
&\quad \times \varepsilon_y^{k+\ell} (\varepsilon_x\varepsilon_{\xbar}\varepsilon_y-1) \phi_{r_{M+1}r_{M+2}\rbar_{\Mbar+1}\rbar_{\Mbar+2}\widehat{s_k}\widehat{s_{\ell}}}\Bigl((x^{4M},y_k,y_{\ell}), (\xbar^{4\Mbar},y_k,y_{\ell}), \bigl(y^{4N} \setminus \{y_k, y_{\ell}\}\bigr)\Bigr),
\end{align}
assuming $k<\ell$. This is a rather unwieldy expression, but the only aspect that matters at this point is the factor $\varepsilon_x\varepsilon_{\xbar}\varepsilon_y-1$, which makes the whole expression vanish if \eqref{productepsilons} holds. 
Here are all six commutators, with those terms containing a factor of $\varepsilon_x\varepsilon_{\xbar}\varepsilon_y-1$ left out:
\begin{subequations}
\begin{align}
&\Big[i\partial_{x_i^0} - H_{x_i}, i\partial_{x_j^0} - H_{x_j}\Big]  = 0\,,\\
&\Big[i\partial_{\xbar_{\ibar}^0} - H_{\xbar_{\ibar}}, i\partial_{\xbar_{\jbar}^0} - H_{\xbar_{\jbar}}\Big] = 0\,,\\
&\Big[i\partial_{y_k^0} - H_{y_k}, i\partial_{y_{\ell}^0} - H_{y_{\ell}}\Big] = 0\,,\\
&\Big[i\partial_{x_i^0} - H_{x_i}, i\partial_{\xbar_{\jbar}^0} - H_{\xbar_{\jbar}}\Big] \phi \nonumber \\
&\quad= -(1-\kappa) \,\sqrt{\frac{N+1}{M\Mbar}}\, \varepsilon_x^{i+1} \, \varepsilon_{\xbar}^{\jbar+1} \, \varepsilon_y^N \sum_{s_{N+1}=1}^4 \bigg\{ \big( i\partial_{x_i^0}-H_{x_i}^\free \big) G_{r_i\,\rbar_{\jbar}\,s_{N+1}}(x_i-\xbar_{\jbar}) \bigg\}~\times \nonumber\\
&\quad\quad\quad\times~ \phi_{\widehat{r_i}\, \widehat{\rbar_{\jbar}}\, s_{N+1}}\Bigl(x^{4M} \setminus x_i,\xbar^{4\Mbar} \setminus \xbar_{\jbar},(y^{4N}, \xbar_{\jbar})\Bigr) \nonumber \\
&\quad\quad +\kappa \,\sqrt{\frac{N+1}{M\Mbar}} \, \varepsilon_x^{i+1} \, \varepsilon_{\xbar}^{\jbar+1}\, \varepsilon_y^N \sum_{s_{N+1}=1}^4 \bigg\{ \big( i\partial_{\xbar_{\jbar}^0}-H_{\xbar_{\jbar}}^\free \big) \Gbar_{r_i\,\rbar_{\jbar}\,s_{N+1}}(\xbar_{\jbar}-x_i) \bigg\} ~\times\nonumber\\
&\quad\quad\quad \times~ \phi_{\widehat{r_i}\, \widehat{\rbar_{\jbar}}\, s_{N+1}}\Bigl(x^{4M} \setminus x_i,\xbar^{4\Mbar} \setminus \xbar_{\jbar},(y^{4N}, x_i)\Bigr)\,,\\
\intertext{which vanishes because each curly bracket vanishes due to \eqref{Gevol} and \eqref{Gbarevol}, and}
&\Big[i\partial_{x_j^0} - H_{x_j}, i\partial_{y_k^0} - H_{y_k}\Big] \phi \nonumber \\
&\quad= -\kappa \, \varepsilon_x^{M+j+1}\,  \varepsilon_y^{N+k} \!\!\!\! \sum_{r_{M+1},\rbar_{\Mbar+1},s_{N+1}=1}^4 \!\!\!\! g_{r_{M+1}\rbar_{\Mbar+1}s_k}^* \Gbar_{r_j\rbar_{\Mbar+1}s_{N+1}}(y_k-x_j) ~\times \nonumber \\
&\quad\quad \quad \times ~ \phi_{r_{M+1}\widehat{r_j}\widehat{s_k}s_{N+1}}\Bigl((x^{4M} \setminus x_j,y_k), \xbar^{4\Mbar}, (y^{4N} \setminus y_k, x_j)\Bigr)\,,
\intertext{which vanishes on spacelike configurations for which $x_j$ does not collide with $y_k$ because then $y_k$ is spacelike separated from $x_j$, and it is known \cite[Thm.~1.2 on p.~15]{thaller:1992} that the Green function $\Gbar$ for the Dirac equation vanishes on spacelike vectors. Finally,}
&\Big[i\partial_{\xbar_{\jbar}^0} - H_{\xbar_{\jbar}}, i\partial_{y_k^0} - H_{y_k}\Big] \phi  \nonumber \\
&\quad= -(1-\kappa)\, \varepsilon_{\xbar}^{\Mbar+\jbar+1} \, \varepsilon_y^{N+k} \!\!\!\! \sum_{r_{M+1},\rbar_{\Mbar+1},s_{N+1}=1}^4 \!\!\!\! g_{r_{M+1}\rbar_{\Mbar+1}s_k}^* \, G_{r_{M+1}\rbar_{\jbar}\,s_{N+1}}(y_k-\xbar_{\jbar})~\times \nonumber\\
&\quad\quad \quad \times ~ 
 \phi_{\rbar_{\Mbar+1}\widehat{\rbar_{\jbar}}\,\widehat{s_k}s_{N+1}}\Bigl(x^{4M}, (\xbar^{4\Mbar} \setminus \xbar_{\jbar},y_k), (y^{4N} \setminus y_k, \xbar_{\jbar})\Bigr)\,,
\end{align}
\end{subequations}
which vanishes on spacelike configurations for which $\xbar_{\jbar}$ does not collide with $y_k$ because the Green function $G$ vanishes on spacelike vectors. This completes our consistency proof.

Conversely, consistency on $\sS_{x\xbar y}$ fails if \eqref{productepsilons} is violated. Indeed, in this case, i.e., if $\varepsilon_x\varepsilon_{\xbar}\varepsilon_y=-1$, \eqref{comm3_neu} shows that the commutator (as an operator acting on arbitrary functions $\phi$ on $\sS_{x\xbar y}$) is non-zero at every configuration. By the results of Section~5.5 of \cite{pt:2013c}, this fact at any collision-free configuration implies that the Equations \eqref{multi123} are inconsistent.

We further remark that if, in the reaction $a \rightleftarrows b+c$, some of the particle species coincide (as in $x\rightleftarrows x+y$, $x\rightleftarrows y+y$, or $x\rightleftarrows x+x$), then an appropriate variant of the system \eqref{multi123} can straightforwardly be formulated, and that variant is consistent if and only if $\varepsilon_a \varepsilon_b \varepsilon_c =1$, i.e., if an even number of the three particles are fermions.

\section{Behavior Under Lorentz Transformations}
\label{sec:lorentz}
Our discussion in this section follows the one in Section 2.4 of \cite{pt:2013c} for the multi-time equations considered there. To investigate the behavior of the system \eqref{multi123} under Lorentz transformations, let us first slightly reformulate \eqref{multi123}. Let us write $\partial_{j\mu}$ [respectively, $\partial_{\jbar \mu},\partial_{k\mu}$] for $\partial/\partial x^\mu_j$ [respectively, $\partial/\partial \xbar_{\jbar}^\mu,\partial/\partial y^\mu_k$], so that the name of the particle label conveys whether we are considering an $x$-, $\xbar$-, or $y$-particle. Similarly, let $\gamma_j^\mu$ [$\gamma_{\jbar}^\mu,\gamma_k^\mu$] be the Dirac gamma matrices acting on the index $r_j$ [$\rbar_{\jbar},s_k$]. Starting from \eqref{multi1} [\eqref{multi2},\eqref{multi3}], 
moving the free Hamiltonian to the left-hand side, and multiplying by $\gamma_j^0$ [$\gamma_{\jbar}^0,\gamma_k^0$], we obtain
\begin{subequations}\label{multi456}
\begin{align}
&\Bigl(i\gamma_{j}^\mu\partial_{j\mu} - m_x\Bigr) \phi\bigl(x^{4M},\xbar^{4\Mbar},y^{4N}\bigr) = 
\kappa \sqrt{\frac{N+1}{M\Mbar}} \sum_{\jbar=1}^{\Mbar} \varepsilon_x^{j+1}\,\varepsilon_{\xbar}^{\jbar+1}\,\varepsilon_y^N \,\times\nonumber\\
&\qquad \times   \sum_{s_{N+1}=1}^4 
 \tilde{\Gbar}_{r_j\,\rbar_{\jbar}}^{~~~~\:s_{N+1}}(\xbar_{\jbar}-x_j)\:\: 
\phi_{\widehat{r_j}\,\widehat{\rbar_{\jbar}}\,s_{N+1}}\Bigl(x^{4M}\setminus x_j,\xbar^{4\Mbar}\setminus \xbar_{\jbar}, (y^{4N}, x_j)\Bigr)\label{multi4}\\[3mm]
&\Bigl(i\gamma_{\jbar}^\mu\partial_{\jbar\mu} - m_{\xbar}\Bigr)\phi\bigl(x^{4M},\xbar^{4\Mbar},y^{4N}\bigr) =  (1-\kappa) \sqrt{\frac{N+1}{M\Mbar}}\sum_{j=1}^M \varepsilon_x^{j+1}\,\varepsilon_{\xbar}^{\jbar+1}\,\varepsilon_y^N\, \times\nonumber\\
&\qquad\times \: \sum_{s_{N+1}=1}^4 
\tilde{G}_{r_j \, \rbar_{\jbar}}^{~~~~\: s_{N+1}}(x_j-\xbar_{\jbar})\:\: 
\phi_{\widehat{r_j}\,\widehat{\rbar_{\jbar}}\,s_{N+1}}\Bigl(x^{4M}\setminus x_j, \xbar^{4\Mbar}\setminus \xbar_{\jbar}, (y^{4N}, \xbar_{\jbar})\Bigr)\label{multi5}\\[3mm]
&\Bigl( i\gamma_k^\mu\partial_{k\mu} - m_y\Bigr) \phi\bigl(x^{4M},\xbar^{4\Mbar},y^{4N}\bigr) = 
\sqrt{\frac{(M+1)(\Mbar+1)}{N}} \varepsilon_x^M\,\varepsilon_{\xbar}^{\Mbar}\,\varepsilon_y^{k+1} \,\times\nonumber\\
&\qquad\times \: \sum_{r_{M+1},\rbar_{\Mbar+1}=1}^4
(\tilde{g}^+)^{r_{M+1} \rbar_{\Mbar+1}}_{~~~~~~~~~~~s_k} \:\: 
\phi_{r_{M+1} \,\rbar_{\Mbar+1} \widehat{s_k}}\Bigl( (x^{4M}, y_k),(\xbar^{4\Mbar}, y_k),y^{4N}\setminus y_k \Bigr)\label{multi6}
\end{align}
\end{subequations}
with implicit summation over $\mu$ but not over $j$ or $\jbar$ or $k$, and
\begin{subequations}\label{tildeGthingsdef}
\begin{align}
\tilde{\Gbar}_{r\rbar}^{~~\:s}(\xbar)&=\sum_{r'=1}^4 (\gamma^0)_{rr'} \Gbar_{r'\rbar s}(\xbar)\label{tildeGbardef}\\
\tilde{G}_{r\rbar}^{~~\:s}(x)&=\sum_{\rbar'=1}^4 (\gamma^0)_{\rbar\rbar'} G_{r\rbar' s}(x)\label{tildeGdef}\\
(\tilde{g}^+)^{r\rbar}_{~~s}&=\sum_{s'=1}^4 (\gamma^0)_{ss'} \, g^*_{r\rbar s'}\,.\label{tildeg+def}
\end{align}
\end{subequations}
In \eqref{multi456} and the left-hand side of \eqref{tildeGthingsdef}, upper spin indices refer to $S^*$, the dual space of $S$, while lower ones refer to $S$. The Lorentz-invariant operation $^+$ is defined in Section 2.4 of \cite{pt:2013c}.

The system of equations \eqref{multi456} is Lorentz-invariant if we regard $\tilde{\Gbar},\tilde{G}$ as functions $\sM\to S\otimes S \otimes S^*$, where $\sM$ denotes Minkowski space-time, and $\tilde{g}^+$ as an element of $S^*\otimes S^* \otimes S$. 

In fact, $\tilde{\Gbar}_{r\rbar}^{~~\:s}(\xbar)$ is the Green function (in the variable $\xbar$ and the spin index $\rbar$) with initial spinor covariantly characterized by $\tilde{g}$. Likewise, $\tilde{G}_{r\rbar}^{~~\:s}(x)$ is the Green function (in the variable $x$ and the spin index $r$) with initial spinor covariantly characterized by $\tilde{g}$. That is, the objects $\tilde{\Gbar},\tilde{G}$ and $\tilde{g}^+$ can all be obtained in a unique and covariant way (and with the right transformation behavior) once an element $\tilde{g}\in S\otimes S\otimes S^*$ has been chosen.

\section{Generalization to Curved Space-Time}
\label{sec:curved}

In Section 2.5 of \cite{pt:2013c}, we had outlined the rather straightforward way how the multi-time equations presented there can be generalized to a curved space-time $(\sM,g)$. The multi-time system \eqref{multi123} or, equivalently, \eqref{multi456} can be adapted to curved space-time in the same way. Here is a brief overview. In this setting, $\phi$ is defined on the spacelike configurations in $\Gamma(\sM)^3$ with values
\begin{multline}
\phi(x_1,\ldots,x_M,\xbar_1,\ldots,\xbar_{\Mbar},y_1,\ldots,y_N)\in\\
 S_{x_1}\otimes \cdots \otimes S_{x_M}\otimes S_{\xbar_1} \otimes \cdots \otimes S_{\xbar_{\Mbar}}\otimes S_{y_1}\otimes \cdots \otimes S_{y_N}\,,
\end{multline}
where $S_x$ is a fiber of the bundle $S$ of spin spaces, a vector bundle over the base manifold $\sM$. This can be expressed using the notation $A\boxtimes B$ for the vector bundle over the base manifold $\sA\times \sB$ (obtained from vector bundles $A$ over $\sA$ and $B$ over $\sB$) whose fiber at $(a,b)\in\sA\times\sB$ is
\be
(A\boxtimes B)_{(a,b)} = A_a\otimes B_b\,,
\ee
and correspondingly $A^{\boxtimes n}$ for $A\boxtimes A \boxtimes \cdots \boxtimes A$ with $n$ factors. Then the $(M,\Mbar,N)$-particle sector of $\phi$ is a cross-section of the vector bundle $S^{(M,\Mbar,N)}=S^{\boxtimes M}\boxtimes S^{\boxtimes \Mbar}\boxtimes S^{\boxtimes N}$ over $\sM^{M+\Mbar+N}$. The equations \eqref{multi456} need only minor changes and re-interpretation of symbols, such as that $\partial_{j\mu}$ is now the covariant derivative on $S$ corresponding to the connection naturally associated with the metric of $\sM$, and correspondingly on $S^{\boxtimes (M+\Mbar+N)}$. The coupling matrix $\tilde{g}_{r\rbar}^{~~s}$ gets replaced in \eqref{multi6} by a cross-section $\tilde{g}_{r\rbar}^{~~s}(y_k)$ of the bundle $S\otimes S \otimes S^*$. Similarly, $\tilde\Gbar(\xbar-x)$ in \eqref{multi4} gets replaced by $\tilde\Gbar(\xbar,x)$, which is the appropriate Green function, namely the solution of the free Dirac equation in $\xbar$ with initial spinor at $x$ covariantly characterized by $\tilde{g}(x)$; likewise, $\tilde{G}(x-\xbar)$ in \eqref{multi5} gets replaced by the appropriate Green function $\tilde{G}(x,\xbar)$. 

Our consistency proof still applies.

\bigskip

\noindent{\it Acknowledgments.} 
S.P.\ acknowledges support from Cusanuswerk, from the German--American Fulbright Commission, and from the European Cooperation in Science and Technology (COST action MP1006). R.T.\ acknowledges support from the John Templeton Foundation (grant no.\ 37433) and from the Trustees Research Fellowship Program at Rutgers. 

\end{document}